\documentclass[3p]{elsarticle}

\usepackage{amssymb,amsmath,amsfonts}
\usepackage{verbatim,a4wide}
\usepackage[cp1250]{inputenc}
\usepackage[OT4]{fontenc}
\usepackage[english]{babel}

\usepackage{bm}
\usepackage{graphicx}

\usepackage{epstopdf}
\usepackage{caption}
\usepackage{subfig}
\usepackage{booktabs}
\usepackage{multirow}
\usepackage{hyperref}
\usepackage{color}

\biboptions{square,numbers,sort&compress}

\numberwithin{equation}{section}
\newtheorem{thm}{Theorem}[section]
\newtheorem{lem}[thm]{Lemma}

\newtheorem{alg}[thm]{Algorithm}
\newdefinition{exmp}[thm]{Example}
\newtheorem{prob}[thm]{Problem}
\newdefinition{rem}[thm]{\it Remark }
\newproof{pf}{Proof}
\newcommand{\pih}{\mbox{$\bm p_i^h$}}
\newcommand{\qih}{\mbox{$\bm q_i^h$}}
\newcommand{\lam}{\mbox{$\bm{\lambda}$}}
\newcommand{\Lam}{\mbox{$\bm{\Lambda}$}}

\newcommand{\R}{\mathbb R}

\newcommand{\ip}[2]{\mbox{$\left\langle{#1},{#2}\right\rangle$}}
\newcommand{\der}{\mbox{${\rm\,d}$}}
\newcommand{\dt}{\mbox{${\rm\,d}t$}}
\newcommand{\du}{\mbox{${\rm\,d}u$}}

\newcommand{\Span}{\textrm{\,span}}
\newcommand{\rbinom}[2]{\mbox{$\displaystyle\binom{#1}{#2}^{\!\!-1}\!\!$}}

\makeatletter
\def\ps@pprintTitle{%
  \let\@oddhead\@empty
  \let\@evenhead\@empty
  \let\@oddfoot\@empty
  \let\@evenfoot\@oddfoot
}
\makeatother

\begin{document}
\thispagestyle{empty}

\begin{frontmatter}

	\title{Degree reduction of composite B\'ezier curves}

\author{Przemys{\l}aw Gospodarczyk\corref{cor}}
\ead{pgo@ii.uni.wroc.pl}
\author{Stanis{\l}aw Lewanowicz}
\ead{Stanislaw.Lewanowicz@ii.uni.wroc.pl}
\author{Pawe{\l} Wo\'{z}ny}
\ead{Pawel.Wozny@ii.uni.wroc.pl}
\cortext[cor]{Corresponding author. Fax {+}48 71 3757801}
\address{Institute of Computer Science, University of Wroc{\l}aw,
         ul.~Joliot-Curie 15, 50-383 Wroc{\l}aw, Poland}

\begin{abstract}
	This paper deals with the problem of multi-degree reduction of a composite B\'ezier curve
	with the parametric continuity constraints at the endpoints of the segments.
We present a novel method which is based on the idea of using constrained dual Bernstein polynomials
to compute the control points of the reduced composite curve.
In contrast to other methods, ours minimizes the $L_2$-error for the whole composite curve instead of minimizing
the $L_2$-errors for each segment separately.  As a result, an additional optimization is possible. Examples show that the new method gives
much better results than multiple application of the degree reduction of a single B\'ezier curve.
\end{abstract}


\end{frontmatter}


\section{Introduction}
                                                        \label{S:Intr}
In recent years, many methods have been used to reduce the degree of B\'ezier curves with constraints (see, e.g., \cite{Ahn03,ALPY04,HB16,BSB96,CW02,Eck95,Gos15,GLW16,LPY02,Lu13,Lu15a,LWa06,
LWa08,RLY06,RM11,RM13,WL09}).
Most of  these papers give  methods
of multi-degree reduction of a \textit{single} B\'ezier curve with constrains of endpoints (parametric or geometric) continuity
of arbitrary order with respect to $L_2$-norm.
Observe, however, that degree reduction schemes often need to be combined with the subdivision
algorithm, i.e., a high degree curve is replaced by a number of lower degree curve segments,
or a \textit{composite} B\'ezier curve, and
continuity between adjacent lower degree curve segments should be maintained. Intuitively,
a possible approach in such a case is applying the multi-degree reduction procedure  to one segment of the  curve after another with properly chosen endpoints  continuity constraints. However, in general,
the obtained solution does not minimize the distance between two composite curves.

In this paper, we give the optimal least-squares solution of multi-degree reduction of a composite B\'ezier curve
with the parametric continuity constraints at the endpoints of the segments.
More specifically, we consider the following approximation problem.
\begin{prob} [\sf $C^{\bm r}$-constrained multi-degree reduction  of the composite B\'ezier curve]
                                                \label{P:main}

Let $a=t_0<t_1<\ldots<t_s=b$ be a partition of the interval $[a,\,b]$.
Let there be given a degree $\bm n=(n_1,n_2,\ldots,n_s)$ composite B\'ezier curve $P(t)$  ($t\in[a,\,b]$)
in $\R^d$ that in the interval $[t_{i-1},\,t_i]$
($i=1,2,\ldots,s$) is exactly represented as a~B\'ezier curve $P_i(u)$ ($u\in[0,\,1]$) of degree $n_i$, i.e.,
\begin{equation*}
	\label{E:P}
	P(t)=P_i(u):=\sum_{j=0}^{n_i}p_{i,j}\,B^{n_i}_j(u)
	\qquad (t_{i-1}\le t\le t_i;\;u:=(t-t_{i-1})/{h_i}),
\end{equation*}
where $h_i:=t_i-t_{i-1}$, and
\[
   B^n_{j}(u):=\binom nj u^j(1-u)^{n-j} \qquad (0\le j\le n)
\]
are the {Bernstein basis polynomials} of degree $n$.
Find a   composite B\'ezier curve $Q(t)$  ($t\in[a,\,b]$) of degree $\bm m=(m_1,m_2,\ldots,m_s)$
which in the interval $[t_{i-1},\,t_i]$ ($i=1,2,\ldots,s$)
is exactly represented as a B\'ezier curve $Q_i(u)$ ($u\in[0,\,1]$) of degree $m_i<n_i$, i.e.,
\begin{equation}
	\label{E:Q}
	Q(t)=Q_i(u):=\sum_{j=0}^{m_i}q_{i,j}\,B^{m_i}_j(u)
	\qquad (t_{i-1}\le t\le t_i;\;u:=(t-t_{i-1})/{h_i}),
\end{equation}
such that the squared $L_2$-error
\begin{equation}
	\label{E:E}
	E:=\int_{a}^{b}\|P(t)-Q(t)\|^2\dt=\sum_{i=1}^{s}E_i,	
	\end{equation}
where
\begin{equation*}
	\label{E:Ei}
	E_i:=\int_{t_{i-1}}^{t_i}\|P(t)-Q(t)\|^2\dt=h_{i}\int_{0}^{1}\|P_i(u)-Q_i(u)\|^2\du,		
	\end{equation*}
reaches the minimum under the additional conditions that
\begin{align}
	\label{E:constr-left}
	&\dfrac{\der^j Q(t)}{\dt^j}\Big|_{t=t_0}=\dfrac{\der^j P(t)}{\dt^j}\Big|_{t=t_0}
	\qquad (j=0,1,\ldots,r_0),\\[2ex]
	\label{E:constr-mid}
	&\dfrac{\der^j Q(t)}{\dt^j}\Big|_{t=t_i-}=\dfrac{\der^j Q(t)}{\dt^j}\Big|_{t=t_i+}\qquad (i=1,2,\ldots,s-1;\;j=0,1,\ldots,r_i),\\[2ex]
	 \label{E:constr-right}
	 &\dfrac{\der^j Q(t)}{\dt^j}\Big|_{t=t_s}=\dfrac{\der^j P(t)}{\dt^j}\Big|_{t=t_s}
	\qquad (j=0,1,\ldots,r_s),
	 \end{align}
 where $r_j\ge0$ ($j=0,1,\ldots,s$) and $r_{i-1}+r_{i}< m_i-1$ ($i=1,2,\ldots,s$).
We will say that the  curves $P$ and $Q$ satisfy the $C^{\bm r}-$\emph{continuity conditions}
	at the points $t_0,t_1,\ldots,t_s$, where we use the notation $\bm r:=(r_0,r_1,\ldots,r_s)$.
Here $\|\cdot\|$ is the Euclidean vector norm.
\end{prob}

\begin{rem}\label{R:OldNew}
	One may think that the conditions
$$
	\dfrac{\der^j Q(t)}{\dt^j}\Big|_{t=t_i}=\dfrac{\der^j P(t)}{\dt^j}\Big|_{t=t_i}
	\qquad (i=1,2,\ldots,s-1;\;j=0,1,\ldots,r_i)
$$
would be more natural than \eqref{E:constr-mid}.
However, in contrast to our new method, such an approach leaves no room for additional optimization.
\end{rem}

\begin{rem}\label{R:ProbVar}
	Sometimes, it may be useful to interpolate the endpoints of the original segments (see Example~\ref{Ex:L}), i.e., to demand that $Q(t_i)= P(t_i)$ holds for  $i=1,2,\ldots,s-1$. In such a case, constraints \eqref{E:constr-mid} should be appropriately modified by restricting the range of $j$ to $1,2,\ldots,r_i$.
\end{rem}

The paper is organized as follows. In Section~\ref{S:Prelim}, we recall some results which are later applied in
the solution of the problem,  given in Section~\ref{S:Main}.
Several illustrative examples are presented in Section~\ref{S:Examples}. Finally, Section~\ref{S:Conc} contains some
concluding remarks.

We end this section with introducing some notation. The \textit{shifted factorial} is defined by $(c)_0:=1$,
$(c)_j:=c(c+1)\cdots (c+j-1)$\ ($j=1,2,\ldots$).
The  \textit{iterated forward difference operator} $\Delta^j$ is given by
\[
	\Delta^j \alpha_k := \Delta^{j-1}\alpha_{k+1} - \Delta^{j-1}\alpha_{k} \quad(j =1,2,\ldots)
	\quad \mbox{and}\quad 	\Delta^0\alpha_k := \alpha_k.
\]
Moreover, we adopt the convention that in an expression of the form $\Delta^j \gamma_{i,k}$
the operator $\Delta^j$ acts on the second variable (first variable being fixed), e.g.,
\[
		\Delta^2 \gamma_{i,k}=\Delta \gamma_{i,k+1}-\Delta \gamma_{i,k}
		= \gamma_{i,k+2}-2 \gamma_{i,k+1}+ \gamma_{i,k}.
\]

\section{Preliminaries}
			\label{S:Prelim}

	Let us denote by $\Pi_m^{(k,l)}=\Span\{B^m_{k+1},B^m_{k+2},\ldots,B^m_{m-l-1}\}$,
where $k$ and $l$ are natural numbers such that $k+l < m-1$,
 the space of all polynomials of  degree at most $m$, whose derivatives of order $\le k$ at $t=0$
 and of order $\le l$ at $t=1$ vanish.
There is a unique \textit{dual constrained Bernstein basis of degree} $m$ (see, e.g., \cite{Jue98}),
$D^{(m,k,l)}_{k+1},D^{(m,k,l)}_{k+2},\ldots,D^{(m,k,l)}_{m-l-1}$,
satisfying
$\ip{D^{(m,k,l)}_j}{B^m_h}=\delta_{j,h}$ ($j,h=k+1,k+2,\ldots,m-l-1$),
where the inner product $\ip\cdot\cdot$ is given by
\(
	\ip fg :=\int_{0}^{1} f(t)g(t)\dt,
\)
$\delta_{j,h}$ equals $1$ if $j=h$, and $0$ otherwise.
The table of the  coefficients $C_{j,h}\equiv C_{j,h}(m,k,l)$  in the connection formulas
\begin{equation}
	\label{E:constrDinB}
	D^{(m,k,l)}_j(x)=\sum_{h=k+1}^{m-l-1}C_{j,h}(m,k,l)\,  B^{m}_h(x)
	\qquad (j=k+1,k+2,\ldots,m-l-1)
\end{equation}
can be efficiently computed using the following algorithm (by convention,
	$C_{j,h} := 0$ if $j$ or $h\not\in\{k+1,\ldots,m-l-1\}$):
\begin{alg}[\sf{Computing the $C$-table} \cite{LW11}]
        \label{A:Ctab}
\ \begin{enumerate}
\itemsep2pt
\item Compute the entries of  the first row of the table by
	\begin{equation*}	\label{E:C-start}
\begin{array}{l}
C_{k+1,m-l-1}:=\rbinom{m}{k+1}\,\rbinom{m}{l+1}\;
		  \dfrac{(-1)^{m-k-l-2}(m+k+l+3)!}
	              {(m-k-l-2)!(2k+2)!(2l+2)!},\\[3ex]
	              C_{k+1,h}:=\dfrac{(h-m)(h-k)(h+k+3)}{(h+1)[(h-m)^2-(l+1)^2]}\,C_{k+1,h+1}
	              \quad(h=m-l-2,\ldots,k+2,k+1).
\end{array}
\end{equation*}
\item For $j=k+1,k+2,\ldots,m-l-2$, compute
	\begin{align*}\label{E:C-rec}
		C_{j+1,h}&=\frac1{A(j)}\,\left\{2(j-h)(j+h-m)\,C_{j,h}
		+B(h)\,C_{j,h-1} +A(h)\,C_{j,h+1}-B(j)\,C_{j-1,h}\right\}\nonumber \\
	                  &\hphantom{\frac1{A(i)}\,\left\{2(i-j)(i+j-m)\,C_{i,j}+B(j)\,C_{i,j-1}\right.}
	                  (h=k+1,k+2,\ldots, m-l-1)
\end{align*}
with
$A(u):=(u-m)(u-k)(u+k+2)/(u+1)$, $B(u):=u(u-m-l-2)(u-m+l)/(u-m-1)$.
\end{enumerate}
\end{alg}

\begin{rem}\label{R:Ctab-sym}
	Note that the complexity of the algorithm is $O(m^2)$.
	In \cite[Proposition~3]{Lu15a}, Lu observed that\---due to some symmetries in the $C$-table\---%
	only part of the entries must be calculated as in Algorithm~\ref{A:Ctab}.
\end{rem}

In the solution of Problem~\ref{P:main} given in Section~\ref{S:Main}, we need
the following lemma which follows easily from the results obtained in
\cite[Thm 4.1]{LW11} and \cite[Thm 4.1]{WL09}.
  \begin{lem}[\sf Multi-degree reduction of a B\'ezier curve with prescribed boundary control points]
  	  \label{L:modred}
  	  Let there be given natural numbers $n,m,k,l$ such that
  	  $m<n$ and $k+l<m-1$.
  	  Let $P^\ast$ be the B\'ezier curve of degree $n$,
  	  \[ 
	P^\ast(u):=\sum_{j=0}^{n}p_{j}\,B^{n}_j(u)\qquad (0\le u\le 1).
	  \] 
	The B\'ezier curve of degree $m$,
	\[ 
	Q^\ast(u):=\sum_{j=0}^{m}q_{j}\,B^{m}_j(u)
	\qquad (0\le u\le1),
	\] 
having the prescribed control points $q_{0},q_{1},\ldots,q_{k}$ and
$q_{m-l}, q_{m-l+1}$, $\ldots,q_{m}$,
	and the  inner control points
	\begin{align*}
		q_{j}&:=\sum_{h=k+1}^{m-l-1}C_{j,h}(m,k,l)\,v_h
		-\left(\sum_{h=0}^{k}+\sum_{h=m-l}^{m}\right)K_{j,h}\,q_h 		
		\qquad(j=k+1,k+2,\ldots,m-l-1),		
	\end{align*}
	where
	\begin{align*}
		v_h&:=(n+m+1)^{-1}\binom{m}{h}
		\sum_{i=0}^{n}\binom{n}{i}\rbinom{n+m}{i+h}\,p_i,  \\[1ex]		
		K_{j,h}&:=	\binom{m}{h}\rbinom{m}{j}
		\frac{(-1)^{j-k-1}(k+1-h)_{m-k-l-1}(k+2)_h(l+2)_{m-h}}
		{(j-h)(j-k-1)!(m-l-j-1)!(k+2)_j(l+2)_{m-j}},
	\end{align*}
and $C_{j,h}(m,k,l)$ are the coefficients in \eqref{E:constrDinB},
	gives the least value of the squared $L_2$-error
	\[
		E^\ast:= \int_{0}^{1}\|P^\ast(u)-Q^\ast(u)\|^2\du,	
	\] 	
   equal to
   \[
  	  E^\ast= \sum_{h=1}^{d}\left[I_{n,n}(\mathbf{p}^h,\mathbf{p}^h) + I_{m,m}(\mathbf{q}^h,\mathbf{q}^h)
                                  -2I_{n,m}(\mathbf{p}^h,\mathbf{q}^h)\right],
   \]
  where we use the notation $\mathbf{p}^h$, $\mathbf{q}^h$ for the vectors of \emph{h}th coordinates of points $p_0,p_1,\ldots,p_n$ and
  $q_0,q_1,\ldots,q_m$, respectively, and where for $\bm x:=[x_0,x_1,\ldots,x_N]$ and
  $\bm y:=[y_0,y_1,\ldots,y_M]$, we define
  \[
  	  I_{N,M}(\bm x,\bm y):=(N+M+1)^{-1}
  	  \sum_{i=0}^{N}\sum_{j=0}^{M}\rbinom{N+M}{i+j}\binom{N}{i}\binom{M}{j}x_iy_j.
  	  \]	

\end{lem}

\section{Main result}
\label{S:Main}
Recall that for an arbitrary B\'ezier curve of degree $N$,
\[
	W_N(u)=\sum_{h=0}^{N}w_h\,B^N_h(u) \qquad (0\le u\le1),
\]
the following well-known formulas hold (see, e.g., \cite[\S5.3]{Far02}):
\begin{equation*}\label{E:endpointsder}
	\begin{array}{l}
	\displaystyle \dfrac{\der^j W_N(u)}{\du^j}\Big|_{u=0}
	=(N-j+1)_j\Delta^jw_0=
	(N-j+1)_j\sum_{h=0}^{j}(-1)^{j+h}\binom{j}{h}w_{h},\\[3ex]
	\displaystyle \dfrac{\der^j W_N(u)}{\du^j}\Big|_{u=1} =(N-j+1)_j\Delta^j w_{N-j}=
	(N-j+1)_j\sum_{h=0}^{j}(-1)^{j+h}\binom{j}{h}w_{N-j+h}.
	\end{array}
	\end{equation*}
Moreover, notice that
\begin{align*}
	\dfrac{\der^j P(t)}{\dt^j}\Big|_{t=t_0}&=h_1^{-j}\dfrac{\der^j P_1(u)}{\du^j}\Big|_{u=0},  \\[1ex]
	\dfrac{\der^j P(t)}{\dt^j}\Big|_{t=t_s}&=h_s^{-j}\dfrac{\der^j P_s(u)}{\du^j}\Big|_{u=1},  \\[1ex]
	\dfrac{\der^j Q(t)}{\dt^j}\Big|_{t=t_i+}&=h_{i+1}^{-j}\dfrac{\der^j Q_{i+1}(u)}{\du^j}\Big|_{u=0}
        \qquad (i=0,1,\ldots,s-1),\\[1ex]
        \dfrac{\der^j Q(t)}{\dt^j}\Big|_{t=t_i-}&=h_i^{-j}\dfrac{\der^j Q_i(u)}{\du^j}\Big|_{u=1}
        \qquad (i=1,2,\ldots,s),	
	\end{align*}
where the notation used is that of Problem~\ref{P:main}.
Using the above equations in \eqref{E:constr-left} and \eqref{E:constr-right}, we obtain
\begin{align}
	\label{E:q1-first}
	&q_{1,j}=\binom{n_1}{j}\binom{m_1}{j}^{\!-1}\,\Delta^jp_{1,0}-
	\sum_{h=0}^{j-1}(-1)^{j+h}\binom{j}{ h}q_{1,h}\qquad (j=0,1,\ldots,r_0),\\[2ex]
	\label{E:qs-last}
          & \displaystyle q_{s,m_s-j}
          =(-1)^j\binom{n_s}{j}\binom{m_s}{j}^{\!-1}\,\Delta^jp_{s,n_s-j}-
          \sum_{h=1}^{j}(-1)^h\binom{j}{ h}q_{s,m_s-j+h}\qquad (j=0,1,\ldots,r_s).
	  \end{align}
  Similarly,
  conditions \eqref{E:constr-mid} imply that  for $i\in\{1,2,\ldots,s-1\}$, we have
  \begin{equation}
  	  \label{E:qi-mid}
  	 \left. \begin{array}{l}
  	 \displaystyle q_{i,m_i-h}=\kappa_i+\sum_{j=1}^{h}(-1)^j\binom{h}{j}a_{i,j}\lam_{i,j}, \\[2ex]
  	 \displaystyle 	  q_{i+1,h}=\kappa_i+\sum_{j=1}^{h}\binom{h}{j}a_{i+1,j}\lam_{i,j}
  	  	  \end{array}\quad \right\}\quad(h=0,1,\ldots,r_i),
  	  	  \end{equation}
  where
  \begin{equation*}
  	  \label{E:aij}
  	  a_{i,j}:=\frac{h_{i}^j}{(m_i-j+1)_j},
  	  \end{equation*}
  for an arbitrary point  $\kappa_i=(\kappa^1_i,\kappa_i^2,\ldots,\kappa_i^d)$ and vectors $\lam_{i,j}=[\lambda_{i,j}^1,\lambda_{i,j}^2,\ldots,\lambda_{i,j}^d]$. In particular, we have
  \begin{align*}
  	  & \begin{array}{l}
  	  q_{i,m_i}=\kappa_i,\quad q_{i,m_i-1}=\kappa_i-a_{i,1}\lam_{i,1},\\[2ex]
  	  q_{i+1,0}=\kappa_i,\quad q_{i+1,1}=\kappa_i+a_{i+1,1}\lam_{i,1},
  	  \end{array}\\[1ex]
  	  &\begin{array}{l}
  	  q_{i,m_i-2}=\kappa_i-2a_{i,1}\lam_{i,1}+a_{i,2}\lam_{i,2},
  	  \\[2ex]
  	  q_{i+1,2}=\kappa_i+2a_{i+1,1}\lam_{i,1}+a_{i+1,2}\lam_{i,2},
  	  \end{array}\\[1ex]
  	  &\begin{array}{l}
  	  q_{i,m_i-3}=\kappa_i-3a_{i,1}\lam_{i,1}+3a_{i,2}\lam_{i,2}-a_{i,3}\lam_{i,3},
  	  \\[2ex]
  	  q_{i+1,3}=\kappa_i+3a_{i+1,1}\lam_{i,1}+3a_{i+1,2}\lam_{i,2}+a_{i+1,3}\lam_{i,3}.
  	  \end{array}
  	  \end{align*}

Coming back to the problem of constrained multi-degree reduction of a composite B\'ezier curve
(see Problem~\ref{P:main}), let us observe that for any $i\in\{1,2,\ldots,s\}$ the formulas \eqref{E:q1-first}--\eqref{E:qi-mid}
with \textit{fixed parameters}   $\kappa_i$ and $\lam_{i,j}$  constitute constraints
of the form demanded in Lemma~\ref{L:modred}.
Now, by applying this lemma for $i=1,2,\ldots,s$ to $P^\ast:=P_i$, $Q^\ast:=Q_i$ with $n:=n_i$, $m:=m_i$, $k:=r_{i-1}$ and $l:=r_i$, we obtain
the set of the segments of the composite B\'ezier curve $Q$ \eqref{E:Q} with control points depending on the parameters.
In order to solve Problem \ref{P:main}, we have to determine the \textit{optimum values} of the parameters (cf. Remark \ref{R:OldNew}).

Let us denote
\begin{equation*}
	\label{E:Lam}
	\Lam_i:=[ \kappa_{i},\lam_{i,1},\lam_{i,2},\ldots,\lam_{i,r_{i}}]^T=
	[\kappa^1_{i},\ldots,\kappa^d_i;\lambda^1_{i,1},\ldots,\lambda^d_{i,1};
	\ldots;\lambda^1_{i,r_{i}},\ldots,\lambda^d_{i,r_{i}}]^T \in\R^{\rho_i},
	\end{equation*}
where $\rho_i:=(r_i+1)d$.
The \textit{optimum values} of the parameters can be obtained by minimizing the error function
\eqref{E:E},
\begin{align}\label{E:Epar}
	E &\equiv E(\Lam_1;\ldots;\Lam_{s-1})\nonumber \\[1ex]
	&= E_1(\Lam_1) + \sum_{i=2}^{s-1}E_i(\Lam_{i-1};\Lam_i) + E_s(\Lam_{s-1}),
\end{align}
where
\begin{equation}
	\label{E:Ei-expr}
	E_i=h_{i} \sum_{h=1}^{d}\left[I_{n_i,n_i}(\pih,\pih)+I_{m_i,m_i}(\qih,\qih)
  	  -2I_{n_i,m_i}(\pih,\qih)\right],	
  	  \end{equation}
  $E_1 \equiv E_1(\Lam_1)$, $E_i \equiv E_i(\Lam_{i-1};\Lam_i)$ $(i=2,3,\ldots,s-1)$, $E_s \equiv E_s(\Lam_{s-1})$,
  with $\pih$, $\qih$ being the vectors of $h$th coordinates of points
  $p_{i,0},p_{i,1},\ldots,p_{i,n_i}$ and $q_{i,0},q_{i,1},\ldots,q_{i,m_i}$, respectively. For a minimum of $E$, it is necessary that the derivatives of $E$
	with respect to the parameters $\kappa_i^h$ and $\lambda_{i,j}^h$ are zero, which yields a
	system of $\sigma$ \textit{linear} equations with $\sigma$ unknowns, where $\sigma:=\rho_1+\rho_2+\ldots+\rho_{s-1}$.
	Hence, for $i=1,2,\ldots,s-1$, we have
\begin{align}
  \label{E:sys1}
		    \frac{\partial E}{\partial \kappa_{i}^h}&=\frac{\partial} {\partial \kappa_{i}^h}(E_i+E_{i+1})=0\qquad
		(h=1,2,\ldots,d),\\[1ex]
		\label{E:sys2}
		\frac{\partial E}{\partial \lambda_{i,j}^h}&=\frac{\partial} {\partial \lambda_{i,j}^h}(E_i+E_{i+1})=0
		\qquad (j=1,2,\ldots,r_i;\; h=1,2,\ldots,d).
\end{align}

Now, we summarise the whole idea in Algorithm~\ref{A:Alg}. Note that for $i=1,2,\ldots,s$, the use of Lemma~\ref{L:modred} requires computation
of $C$-table (see Algorithm~\ref{A:Ctab}). The entries of $i$th $C$-table are denoted by $C_{j,h}^{(i)}$ $(j,h=r_{i-1}+1,r_{i-1}+2,\ldots,m_i-r_{i}-1)$.

\begin{alg}[\sf $C^{\bm r}$-constrained multi-degree reduction of composite B\'ezier curves]\label{A:Alg}
	\ \\[0.5ex]
	\noindent \texttt{Input}: $s$, $n_i, m_i$, $p_{i,0},p_{i,1},\ldots,p_{i,n_i}$
	($i=1,2,\ldots,s$),\\[0.5ex]	
	\noindent \hphantom{\texttt{Input}:}  $r_0,r_1,\ldots,r_s$, $a=t_0<t_1<\ldots<t_s=b$\\[0.5ex]
\noindent \texttt{Output}: solution $q_{i,j}\ (i=1,2,\ldots,s;\;j=0,1,\ldots,m_i)$ of Problem \ref{P:main}
\begin{description}
\itemsep2pt
\item[\texttt{Step 1}.] For $i=1,2,\ldots,s$, compute $C_{j,h}^{(i)}$ $(j,h=r_{i-1}+1,r_{i-1}+2,\ldots,m_i-r_i-1)$
	using Algorithm~\ref{A:Ctab}, assuming that  $m := m_i$,
	$k := r_{i-1}$, $l := r_i$.
\item[\texttt{Step 2}.] Compute $E_i\ (i=1,2,\ldots,s)$ by \eqref{E:Ei-expr}.
\item[\texttt{Step 3}.] Solve the system of linear equations \eqref{E:sys1}, \eqref{E:sys2}.
\item[\texttt{Step 4}.] Compute $q_{1,j}\ (j=0,1,\ldots,r_0)$ by \eqref{E:q1-first}.
\item[\texttt{Step 5}.] Compute $q_{s,m_s-j}\ (j=0,1,\ldots,r_s)$ by \eqref{E:qs-last}.
\item[\texttt{Step 6}.] Compute $q_{i,m_i-h}$ and $q_{i+1,h} \ (i=1,2,\ldots,s-1;\;h=0,1,\ldots,r_i)$ by \eqref{E:qi-mid}.
\item[\texttt{Step 7}.] For $i=1,2,\ldots,s$, set $P^\ast:=P_i$, $Q^\ast:=Q_i$, $n:=n_i$, $m:=m_i$, $k:=r_{i-1}$, $l:=r_i$, and
compute the control points $q_{i,j}\ (j=r_{i-1}+1,r_{i-1}+2,\ldots,m_i-r_i-1)$ using Lemma~\ref{L:modred}.
\item[\texttt{Step 8}.] Return $q_{i,j}\ (i=1,2,\ldots,s;\;j=0,1,\ldots,m_i)$.
\end{description}
\end{alg}

\begin{rem}\label{R:CopAlg}
	In case, where  interpolation conditions are imposed at the endpoints of the original segments (see Remark~\ref{R:ProbVar}),  Algorithm \ref{A:Alg} must be slightly modified. Note that the parameter
	$\kappa_i$ $(i=1,2,\ldots,s-1)$ is the meeting point of the consecutive segments $Q_i$ and $Q_{i+1}$, i.e., $\kappa_i = Q_i(1)= Q_{i+1}(0)$. Therefore, by setting $\kappa_i := P(t_i)$ (cf. \eqref{E:qi-mid}), and by removing the subsystem \eqref{E:sys1} from the system,
	the goal is easily achieved.
\end{rem}

\section{Examples}
		\label{S:Examples}

This section provides of the application of our algorithm.
We give the squared $L_2$-errors $E_i$ ($i=1,2,\ldots,s$) and $E$ (see  \eqref{E:Ei-expr},
	\eqref{E:Epar}) as well as the maximum errors
\begin{align*}
&E_i^{\infty} := \max_{t \in U_{i}} \|P(t) - Q(t)\| \approx \max_{t \in [t_{i-1},\,t_i]} \|P(t) - Q(t)\|,\\
&E^{\infty} := \max_{1 \le i \le s} E_i^{\infty},
\end{align*}
where
$U_{i} := \left\{t_{i-1}, t_{i-1}+\delta, t_{i-1}+2\delta,\ldots, t_i\right\}$
with $\delta := (t_i-t_{i-1})/500$.

Results of the experiments have been obtained in Maple{\small \texttrademark}13, using $32$-digit arithmetic.
The system of linear equations \eqref{E:sys1}, \eqref{E:sys2} is solved using Maple{\small \texttrademark} \texttt{fsolve} procedure.

\begin{exmp}\label{Ex:S}
	Assuming that $s=4$ and $t_0 = 0,\ t_1 = 0.18,\ t_2 = 0.29,\ t_3 = 0.53,\ t_4 = 1$, the composite curve ``Squirrel'' is formed by four B\'ezier segments of degrees $9$, $12$, $12$, and $12$,
	respectively. Control points are given at \url{http://www.ii.uni.wroc.pl/~pgo/squirrel.txt}.
For the results of degree reduction, see Table \ref{Tab:sqr} as well as the corresponding Figures \ref{Fig:0a} and \ref{Fig:0b}.
This example shows that Algorithm \ref{A:Alg} may result in a lower error than multiple application of \cite[Algorithm 1]{WL09}.
Furthermore, the larger  are $r_i$'s, the bigger are differences in errors because we have more degrees of freedom, i.e.,  parameters $\lam_{i,j}$ (see Remark \ref{R:OldNew}).
\begin{table}[ht!]
\captionsetup{margin=0pt, font={scriptsize}}
\centering
\scalebox{1}{
\begin{tabular}{@{}lcccccccc@{}}
\toprule & & & & \multicolumn{2}{c}{Algorithm \ref{A:Alg}} & &
\multicolumn{2}{c}{\cite[Algorithm 1]{WL09}}
\\ \cmidrule{5-6} \cmidrule{8-9}
Parameters & & $i$ & & $E_{i}$ & $E_i^{\infty}$ & &  $E_{i}$ &  $E_i^{\infty}$
\\ \midrule
\multirow{2}{*}{$\bm m=(7,8,8,9)$} & & $1$ & & $3.40e{-}8$ & $1.64e{-}3$ & & $3.40e{-}8$ & $1.01e{-}3$\\
& & $2$ & & $4.98e{-}7$ & $5.20e{-}3$ & & $7.82e{-}7$ & $5.36e{-}3$\\
\multirow{2}{*}{$\bm r=(2,0,3,2,2)$}& & $3$ & & $1.85e{-}6$ & $8.23e{-}3$ & & $7.58e{-}6$ & $1.13e{-}2$\\
& & $4$ & & $1.28e{-}6$ & $3.02e{-}3$ & & $3.24e{-}6$ & $5.07e{-}3$\\
\midrule
\multirow{2}{*}{Summary} & & & & $E$ & $E^{\infty}$ & &  $E$ &  $E^{\infty}$\\
& & & & $3.66e{-}6$ & $8.23e{-}3$ & & $1.16e{-}5$ & $1.13e{-}2$\\
\midrule
\multirow{2}{*}{$\bm m=(7,8,8,9)$} & & $1$ & & $9.62e{-}8$ & $1.62e{-}3$ & & $1.10e{-}7$ & $1.84e{-}3$\\
& & $2$ & & $7.26e{-}7$ & $6.26e{-}3$ & & $1.14e{-}5$ & $1.95e{-}2$\\
\multirow{2}{*}{$\bm r=(3,0,4,2,3)$}& & $3$ & & $3.28e{-}6$ & $9.80e{-}3$ & & $6.81e{-}4$ & $9.51e{-}2$\\
& & $4$ & & $3.03e{-}6$ & $4.34e{-}3$ & & $9.37e{-}6$ & $8.24e{-}3$\\
\midrule
\multirow{2}{*}{Summary} & & & & $E$ & $E^{\infty}$ & &  $E$ &  $E^{\infty}$\\
& & & & $7.13e{-}6$ & $9.80e{-}3$ & & $7.02e{-}4$ & $9.51e{-}2$\\
\midrule
\end{tabular}}
\caption{Squared $L_2$-errors and maximum errors for degree reduction of the composite B\'{e}zier curve ``Squirrel''.}
\label{Tab:sqr}
\end{table}

\begin{figure}[ht!]
\captionsetup{margin=0pt, font={scriptsize}}
\begin{center}
\setlength{\tabcolsep}{0mm}
\begin{tabular}{c}
\subfloat[]{\label{Fig:0a}\includegraphics[width=0.51\textwidth]{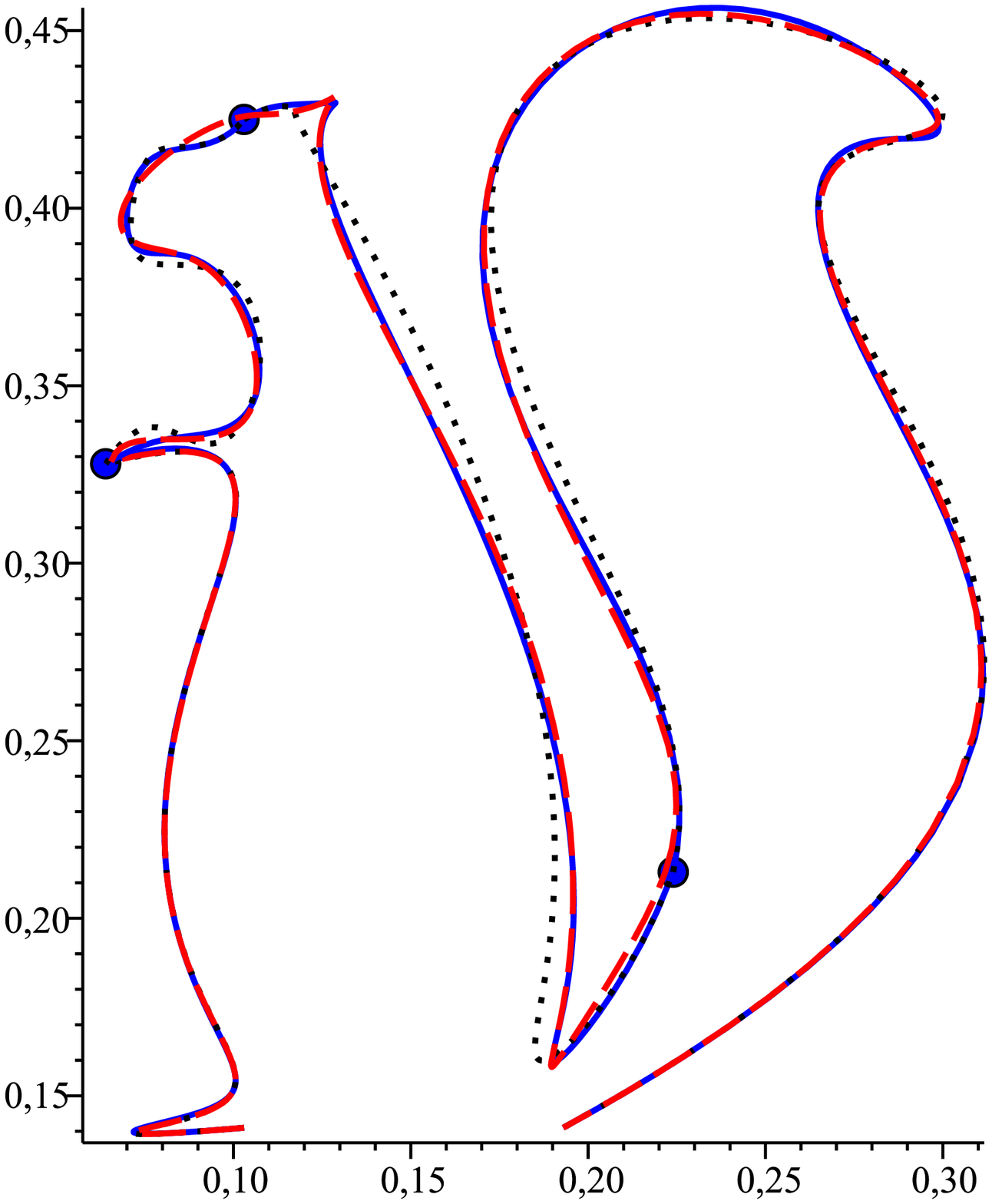}}
\subfloat[]{\label{Fig:0b}\includegraphics[width=0.5\textwidth]{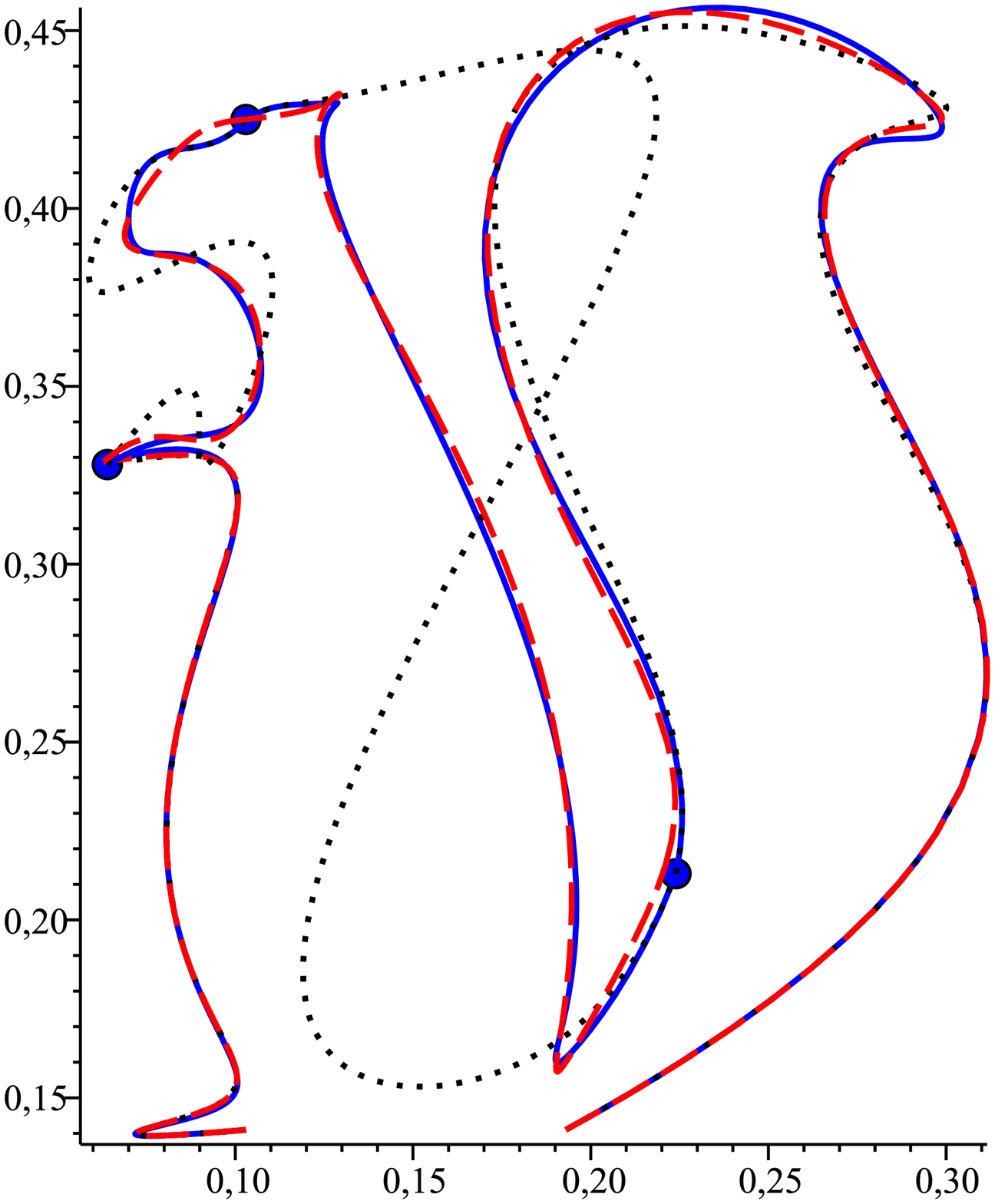}}
\end{tabular}
\caption{Original composite B\'{e}zier curve ``Squirrel'' of degree $\bm n=(9,12,12,12)$ (blue solid line) and degree reduced composite B\'{e}zier curves
    computed using Algorithm \ref{A:Alg} (red dashed line), and \cite[Algorithm 1]{WL09} (black dotted line).
Parameters: (a) $\bm m=(7,8,8,9)$, $\bm r=(2,0,3,2,2)$; and (b) $\bm m=(7,8,8,9)$,
$\bm r=(3,0,4,2,3)$.}
\end{center}
\end{figure}
\end{exmp}

\begin{exmp}\label{Ex:L}
	For $s=2$ and $t_0 = 0,\ t_1 = 0.49,\ t_2 = 1$, we consider the composite curve ``L'' which consists of two B\'ezier segments of degrees $8$ and $12$ having the control points
$\{(0.313, 0.52),$ $(0.198, 0.493),$ $(0.245, 0.412),$ $(0.346, 0.446),$ $(0.466, 0.528),$ $(0.397, 0.518),$ $(0.301, 0.553),$
$(0.296,$ $0.473),$ $(0.299, 0.418)\}$, and
$\{(0.299, 0.418),$ $(0.301, 0.38),$ $(0.323, 0.342),$ $(0.328, 0.294),$ $(0.256,$ $0.252),$ $(0.23, 0.272),$ $(0.173,$ $0.323),$ $(0.237, 0.427),$
$(0.294, 0.278),$ $(0.35, 0.327),$ $(0.4, 0.267),$ $(0.417, 0.296),$ $(0.396, 0.323)\}$, respectively.
The comparison of the results of Algorithm \ref{A:Alg}, algorithm described in Remark \ref{R:CopAlg}, and \cite[Algorithm 1]{WL09} is given in Table \ref{Tab:L} (see also Figure~\ref{Fig:2}). Once again, we observe that the new approach may lead to better results than the older one.
Moreover, we note that in some cases it is useful to interpolate the endpoints of the original segments (see Remarks \ref{R:ProbVar} and \ref{R:CopAlg}).

\begin{figure}[ht!]
\captionsetup{margin=0pt, font={scriptsize}}
\begin{center}
\includegraphics[width=63mm]{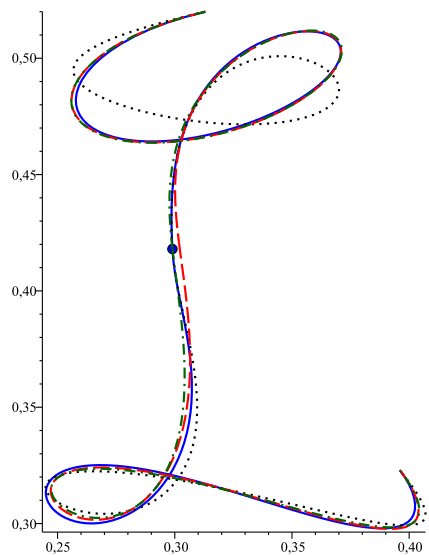}
\caption{Original composite B\'{e}zier curve ``L'' of degree $\bm n=(8,12)$ (blue solid line) and degree reduced composite B\'{e}zier curves
    computed using Algorithm \ref{A:Alg} (red dashed line), algorithm described in Remark \ref{R:CopAlg} (green dash-dotted line), and
    \cite[Algorithm 1]{WL09} (black dotted line). Parameters are specified in Table \ref{Tab:L}.}
    \label{Fig:2}
\end{center}
\end{figure}

\begin{table}[ht!]
\captionsetup{margin=0pt, font={scriptsize}}
\centering
\scalebox{1}{
\begin{tabular}{@{}lccccccccccc@{}}
\toprule & & & & \multicolumn{2}{c}{Algorithm \ref{A:Alg}} & & \multicolumn{2}{c}{Remark \ref{R:CopAlg}} & &
\multicolumn{2}{c}{\cite[Algorithm 1]{WL09}}
\\ \cmidrule{5-6} \cmidrule{8-9} \cmidrule{11-12}
Parameters & & $i$ & & $E_{i}$ & $E_i^{\infty}$ & &  $E_{i}$ &  $E_i^{\infty}$ & &  $E_{i}$ &  $E_i^{\infty}$
\\ \midrule
$\bm m=(6,7)$ & & $1$ & & $1.00e{-}6$ & $3.98e{-}3$ & & $1.23e{-}6$ & $3.10e{-}3$ & & $4.74e{-}5$ & $1.58e{-}2$\\
$\bm r=(1,3,1)$ & & $2$ & & $2.51e{-}6$ & $3.99e{-}3$ & & $4.33e{-}6$ & $5.49e{-}3$ & & $1.91e{-}5$ & $1.08e{-}2$\\
\midrule
\multirow{2}{*}{Summary} & & & &  $E$ & $E^{\infty}$ & &  $E$ &  $E^{\infty}$ & &  $E$ &  $E^{\infty}$\\
& & & & $3.51e{-}6$ & $3.99e{-}3$ & & $5.56e{-}6$ & $5.49e{-}3$ & & $6.65e{-}5$ & $1.58e{-}2$\\
\midrule
\end{tabular}}
\caption{Squared $L_2$-errors and maximum errors for degree reduction of the composite B\'{e}zier curve ``L''.}
\label{Tab:L}
\end{table}
\end{exmp}

\begin{exmp}\label{Ex:G}
Let there be given three B\'ezier curves of degrees $8$, $6$, and $6$, defined by the control points
$\{(0.313, 0.52),$ $(0.198, 0.493),$ $(0.245, 0.412),$ $(0.346, 0.446),$ $(0.466, 0.528),$ $(0.397, 0.518),$ $(0.301, 0.553),$
$(0.296, 0.473),$ $(0.3, 0.422)\}$, $\{(0.305, 0.418),$ $(0.308, 0.344),$ $(0.408,$ $0.342),$ $(0.415,$ $0.445),$ $(0.405, 0.417),$
$(0.402, 0.377),$ $(0.4, 0.365)\}$, and $\{(0.403, 0.36),$ $(0.395,$ $0.249),$ $(0.372, 0.233),$ $(0.225,$ $0.228),$ $(0.302, 0.306),$
$(0.297, 0.308),$ $(0.311, 0.322)\}$, respectively. Note that these B\'ezier curves are not joined (see Figure \ref{Fig:3a}). In spite of that,
we set $t_0 = 0,\ t_1 = 0.45,\ t_2 = 0.68,\ t_3 = 1$, and apply Algorithm \ref{A:Alg} with
$\bm m=(6,5,5)$ and $\bm r=\bm 1:=(1,1,1,1)$. As a result, we obtain the $C^{\bm 1}$-continuous composite B\'ezier curve ``G'' illustrated in Figures \ref{Fig:3b} and \ref{Fig:3c} (errors: $E_1 = 9.94e{-}7$, $E_2 = 2.84e{-}6$, $E_3 = 1.42e{-}6$, $E = 5.25e{-}6$,
     $E_1^{\infty} = 1.06e{-}2$, $E_2^{\infty} = 1.42e{-}2$, $E_3^{\infty} = 9.11e{-}3$, $E^{\infty} = 1.42e{-}2$). In addition, the degrees of the segments are reduced.
 This example shows that our algorithm can serve as a tool for merging of several unconnected B\'ezier curves into a smooth composite B\'ezier curve. Furthermore, in case of $C^0$-continuous input curves, the algorithm can eliminate possible \textit{rough edges} and \textit{corners}.

 \begin{figure}[ht!]
\captionsetup{margin=0pt, font={scriptsize}}
\begin{center}
\setlength{\tabcolsep}{0mm}
\begin{tabular}{c}
\subfloat[]{\label{Fig:3a}\includegraphics[width=0.35\textwidth]{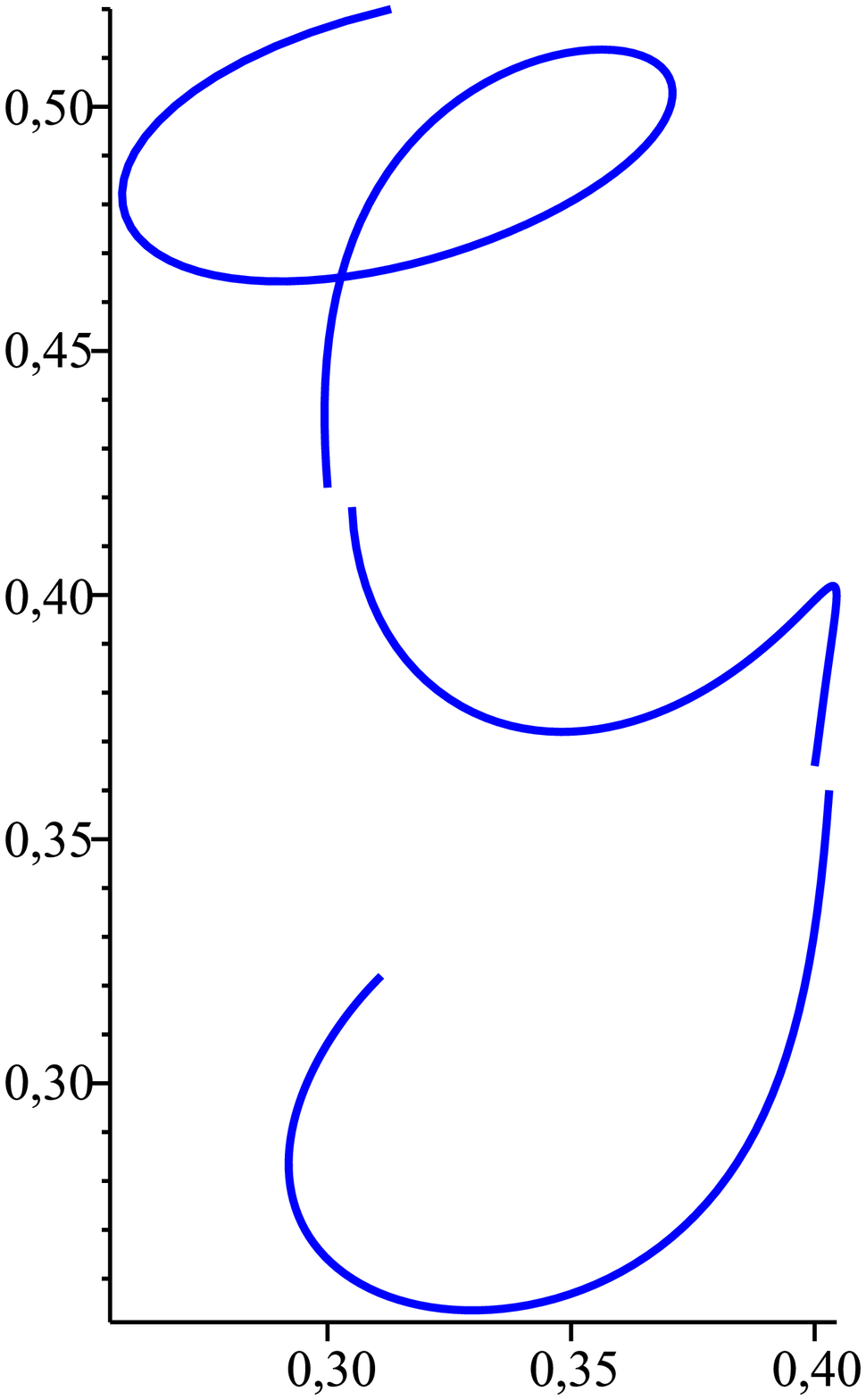}}
\subfloat[]{\label{Fig:3b}\includegraphics[width=0.35\textwidth]{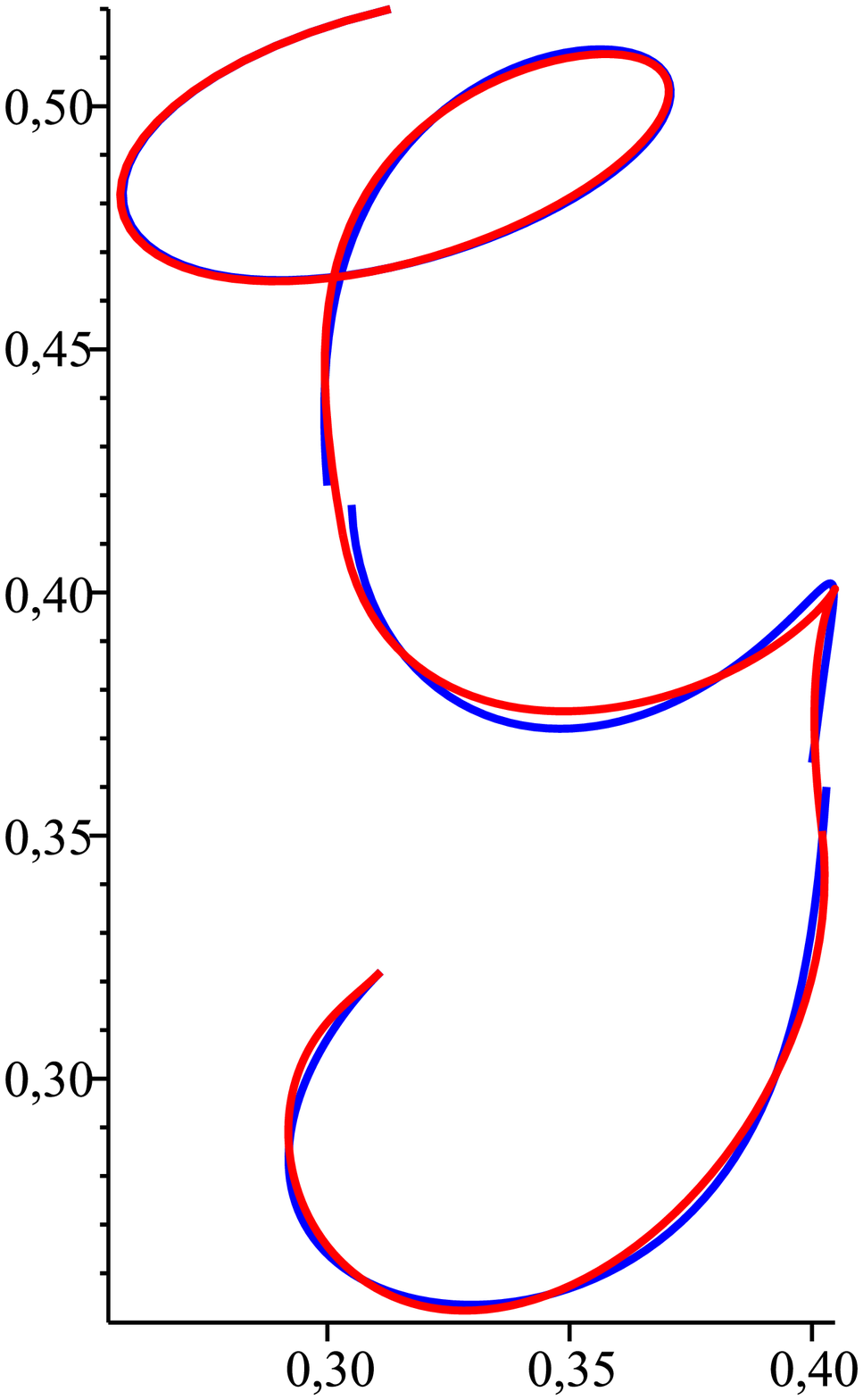}}
\subfloat[]{\label{Fig:3c}\includegraphics[width=0.35\textwidth]{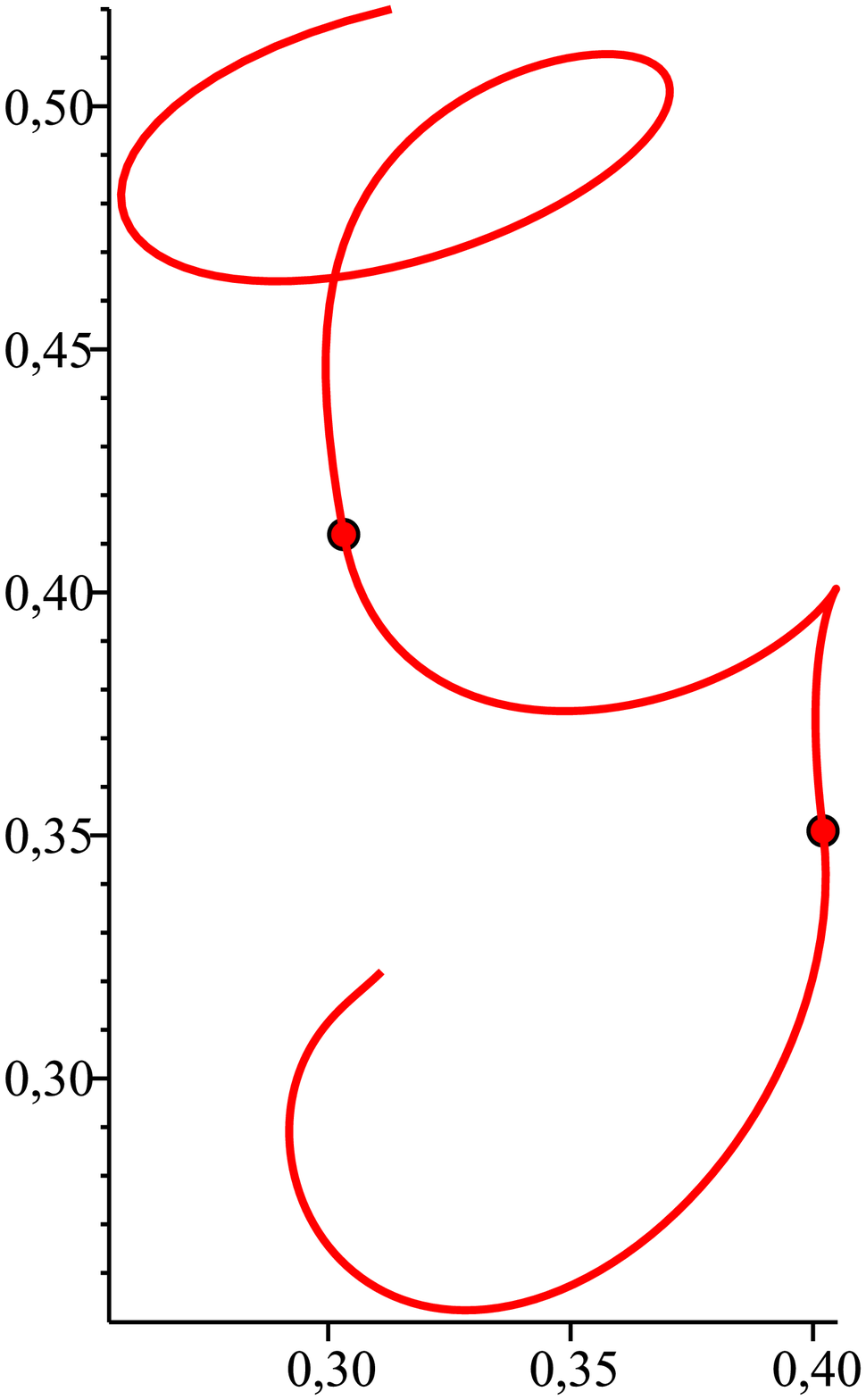}}
\end{tabular}
\caption{Three original B\'{e}zier curves of degrees $8$, $6$, and $6$ (blue solid line) and the $C^{\bm 1}$-continuous composite B\'{e}zier curve ``G'' of degree $\bm m=(6,5,5)$ computed using Algorithm \ref{A:Alg} (red solid line) with $\bm r=\bm 1:=(1,1,1,1)$.}
\end{center}
\end{figure}
\end{exmp}

\section{Conclusions}
		\label{S:Conc}

We propose a novel approach to the problem of multi-degree reduction of composite B\'ezier curves.
In contrast to other methods, ours minimizes the $L_2$-error for the whole composite curve instead of minimizing
the $L_2$-errors for each segment separately. The main idea is to connect consecutive segments of the searched composite curve
using the conditions \eqref{E:constr-mid}. As a result, an additional optimization is possible. The new problem is solved
efficiently using the properties of constrained dual Bernstein polynomial basis.
Examples~\ref{Ex:S} and \ref{Ex:L} show that the new method gives much better results than multiple application of the degree reduction of a single B\'ezier curve. Furthermore, we observe that a slight modification of the method allows for interpolation of endpoints of the original
segments (see Remarks \ref{R:ProbVar} and \ref{R:CopAlg}), which may be useful in some cases (see Example~\ref{Ex:L}).
Moreover, merging of several unconnected B\'ezier curves into a smooth composite B\'ezier curve is also possible (see Example \ref{Ex:G}).

Let us mention that  we have studied also the extended version of Problem~\ref{P:main} where
the parametric continuity conditions \eqref{E:constr-left}--\eqref{E:constr-right} are replaced by
geometric continuity constraints. We have observed that in this case
the error function $E$ becomes a high degree polynomial function of many variables,
even for modest values of $r_i$'s.
Consequently, we have to deal with constrained nonlinear programming problem in order
to find optimal values for the parameters. Experiments show that calculations may be painfully long.
So far, we have not been able to give an efficient algorithm of solving this task.



\begin{thebibliography}{99}
\itemsep 2pt
	
\normalsize	
	
\bibitem{Ahn03}  Y.J. Ahn,  Using Jacobi polynomials for degree reduction of B\'ezier curves
		with $C^k$-constraints, Comput. Aided Geom. Des. 20 (2003), 423--434.
		
\bibitem{ALPY04}  Y.J. Ahn,  B.-G. Lee, Y. Park, J. Yoo,
		Constrained polynomial degree reduction in the $L_2$-norm
		equals best weighted Euclidean approximation of B\'ezier coefficients,
		Comput. Aided Geom. Des. 21 (2004), 181--191.	
	
\bibitem{HB16} R. Ait-Haddou, M. Barto\v{n}, Constrained multi-degree reduction with respect
	to Jacobi norms, Comput. Aided Geom. Des. 42 (2016), 23--30.
	
\bibitem{BSB96} G. Brunnett, T. Schreiber, J. Braun,
		The geometry of optimal degree reduction of B\'ezier curves,
		Comput. Aided Geom. Des. 13 (1996), 773--788.
		
\bibitem{CW02}  G.-D. Chen,  G.-J. Wang,
		Optimal degree reduction of B\'ezier curves with constraints of endpoints continuity,
		Comput. Aided Geom. Des. 19 (2002), 365--377.	
			
\bibitem{Eck95} M. Eck,
		Least squares degree reduction of B\'ezier curves,
		Comput. Aided Des. 27 (1995), 845--851.	
		
\bibitem{Far02} G.E. Farin,
		Curves and Surfaces for Computer-Aided Geometric Design. A Practical Guide, fifth edition,
                Academic Press,  Boston, 2002.

\bibitem{Gos15}  P. Gospodarczyk,
		Degree reduction of B\'{e}zier curves with restricted control points area,
		Comput. Aided Des. 62 (2015), 143--151.

\bibitem{GLW16}  P. Gospodarczyk,  S. Lewanowicz,  P. Wo\'zny,
                $G^{k,l}$-constrained multi-degree reduction of B\'{e}zier curves,
                Numer. Algor. 71 (2016), 121--137.

 \bibitem{Jue98} B. J\"uttler,
 		The dual basis functions for the Bernstein polynomials,
		Adv. Comput. Math. 8 (1998),  345--352.
		
\bibitem{LPY02} B.-G.  Lee, Y. Park, J. Yoo,
		Application of Legendre-Bernstein basis transformations to degree
                elevation and degree reduction,
                Comput. Aided Geom. Des. 19 (2002), 709--718.

\bibitem{LW11}   S. Lewanowicz,  P. Wo\'zny,
                B\'ezier representation of the constrained dual Bernstein polynomials,
                Appl. Math. Comp. 218 (2011), 4580--4586.
                			
\bibitem{Lu13} L. Lu,
                Explicit $G^2$-constrained degree reduction of B\'ezier curves by quadratic optimization,
                J. Comput. Appl. Math. 253 (2013), 80--88.

\bibitem{Lu15a} L. Lu, Gram matrix of Bernstein basis: Properties and applications,
         	J. Comput. Appl. Math. 280 (2015), 37--41.

\bibitem{LWa06} L. Lu, G. Wang,
		Optimal degree reduction of B\'ezier curves with
		$G^2$-continuity, Comput. Aided Geom. Des. 23 (2006), 673--683.
						
\bibitem{LWa08} L. Lu, G. Wang,
		Application of Chebyshev	II-Bernstein basis
		transformations to degree reduction of B\'ezier curves,
		J. Comput. Appl. Math. 221 (2008), 52--65.
				
\bibitem{RLY06} A. Rababah,  B.-G. Lee, J. Yoo,
		A simple matrix form for degree reduction
		of B\'ezier curves using Chebyshev-Bernstein basis transformations,
		Appl. Math. Comp. 181 (2006), 310--318.
		
\bibitem{RM11} A. Rababah, S. Mann,
		Iterative process for ${G}^2$ {m}ulti-degree reduction of  B\'{e}zier curves,
		Appl. Math. Comp. 217 (2011), 8126--8133.		
		
\bibitem{RM13} A. Rababah, S. Mann,
		Linear methods for $G^1$, $G^2$, and $G^3$-multi-degree reduction  of B\'ezier curves,
		Comput. Aided Des. 45 (2013), 405--414.				
		
\bibitem{WL09}  P. Wo\'zny,  S. Lewanowicz,
		Multi-degree reduction of B\'ezier curves
		with constraints, using dual Bernstein basis polynomials,
		Comput. Aided Geom. Des. 26 (2009), 566--579.
	
\end{thebibliography}
\end{document}